\documentclass[aps,prb,reprint,superscriptaddress,floatfix]{revtex4-1}

\usepackage{amsmath,comment}
\usepackage{amsfonts}
\usepackage{amssymb}
\pdfoutput=1
\usepackage{mathtools}
\usepackage{bm}
\usepackage{commath}
\usepackage{bbold}
\usepackage{braket}
\usepackage{hyperref}

\DeclareMathOperator{\sgn}{sgn}
\providecommand{\abs}[1]{\left\vert #1\right\vert}

\renewcommand{\vec}{\mathbf}

\begin{document}
	
\title{Fingerprints of Berry phases in the bulk exciton spectrum of a topological insulator}
\author{Andrew A. Allocca}
 \affiliation{Joint Quantum Institute and Condensed Matter Theory Center, Department of Physics, University of Maryland, College Park,
  Maryland 20742-4111, USA}

\author{Dmitry K. Efimkin}
 \affiliation{The Center for Complex Quantum Systems, The University of Texas at Austin, Austin, Texas 78712-1192, USA}

\author{Victor M. Galitski}
 \affiliation{Joint Quantum Institute and Condensed Matter Theory Center, Department of Physics, University of Maryland, College Park, Maryland 20742-4111, USA}

\begin{abstract}
We examine excitons formed in the bulk of a topological insulator as the system is tuned via a parameter between topological and trivial insulating phases, arguing that nontrivial topology has fingerprints in the spectrum of these excitons. 
The closely related hydrogen atom problem is well known to have a degeneracy due to a hidden symmetry, and the changes to the excitonic spectrum that we find can be understood as a result of breaking of this underlying symmetry due to the Berry phase. 
Furthermore, this phase is found to affect the spectrum in the topological parameter regime much more strongly than in the trivial regime. 
We first construct a semiclassical model of the system to develop qualitative intuition for the effects at play, then move to a more robust numerical simulation of the full quantum system, working with the Bernevig-Hughes-Zhang model of a 2D topological insulator. 
\end{abstract}


\maketitle

\section{Introduction}
Since the discovery of topological insulators (TIs), much work has been done exploring how these new materials can be used to realize exotic new physical phenomena \cite{Hasan2010,Qi2011}. 
One of their defining features is the fact that these materials support robust conducting Dirac states on their surfaces, which themselves have been the focus of a great amount of research.
From the perspective of exciton physics in particular, there have been investigations into the impact that exciton physics at the surface of topological insulators may have on the materials' optical properties \cite{Garate2011,Efimkin2013}, as well as the possibility of exotic interaction effects such as chiral excitons \cite{Efimkin2012} or as a platform for potential realizations of excitonic condensation \cite{Blatt1962, Seradjeh2009}. 
Another work examined the impact of exciton condensation on the quantum spin Hall effect \cite{Pikulin2014}.


Something that has been largely overlooked is how the topological nature of these materials manifests in properties of the bulk. 
Far away from the surface, though global properties are different, the band structure of a topological insulator is qualitatively very similar to that of a trivial insulator or even a semiconductor with a large band gap.
Consequently, optical and transport properties are naively expected to be similar as well, and indeed electrical conductivity through the bulk is exponential small in the size of the gap for both trivial and topological insulators.
Only few studies have been done, however, exploring the effect of nontrivial topological character on other physical phenomena in the bulk.
One such study examined the polarization properties of a 2D topological model, concluding that features of the optical conductivity of this model, including an plasmon resonance absent in graphene or usual 2DEGs, provide a way to identify its topological character via bulk measurements \cite{Juergens2014}.
Another work found that phonon linewidths of bulk optical phonons contain information on band inversions in the electronic spectrum \cite{Saha2015}.
Here we add to this line of inquiry, investigating how the properties of excitons formed from the bulk bands of a topological insulator depend on the topological character.

We consider a model that can be continuously tuned between topologically trivial and nontrivial parameter regimes--the well-studied Bernevig-Hughes-Zhang (BHZ) model \cite{Bernevig2006}, developed to describe the band-inversion physics and resulting topological phase of Hg(Cd)Te quantum wells.
It is given by the Hamiltonian 
\begin{equation}\label{eq:BHZham}
\begin{gathered}
H_\text{BHZ}(\vec{p}) = \begin{pmatrix}
\hat{h}_\vec{p} & 0 \\
0 & \hat{h}_{-\vec{p}}^\ast \\
\end{pmatrix}, \\ 
\hat{h}_\vec{p} = \epsilon_p \hat{\mathbf{1}} + \vec{d}_\vec{p}\cdot \hat{\boldsymbol{\tau}}, \quad \vec{d}_\vec{p} = \left( A p_x, -A p_y, M_\vec{p}\right),
\end{gathered}
\end{equation}
where $\epsilon_p = C - D p^2$ is the electron-hole asymmetry, $M_\vec{p} = M - B p^2$ is the momentum-dependent Dirac mass, $\hat{\boldsymbol{\tau}}$ is the vector of Pauli matrices, $\hat{\mathbf{1}}$ is the unit matrix, and 
$A, B, C, D,$ and $M$ are material parameters. 
The Hamiltonian is invariant under both time reversal and inversion, discussed in Appendix \ref{sec:symmetries}.
Importantly for our purposes, the mass-like parameter $M$ is related to the thickness of the quantum well and can be tuned between positive and negative values.
Changing this sign changes the relative sign of the $p=0$ and $p\to\infty$ limits of the mass term $M_\vec{p}$, corresponding to two topologically distinct phases. 
This model provides the ideal testbed for our analysis.


Important quantities to consider in the context of topological insulators are the Berry connection, the Berry curvature, and the resulting Berry phase.
It has been well established that Berry physics can lead to a shift and splitting of otherwise degenerate exciton energy levels even in a system with trivial topological character \cite{Srivastava2015,Zhou2015}, i.e.\ without the usual hallmarks of topological phases such as protected edge states or a nontrivial topological index.
It is therefore reasonable to expect that similar effects will be seen in the topological phase of the BHZ model, which does display all of these features as a direct result of Berry physics.

\begin{figure}
\center
\includegraphics[width=\columnwidth]{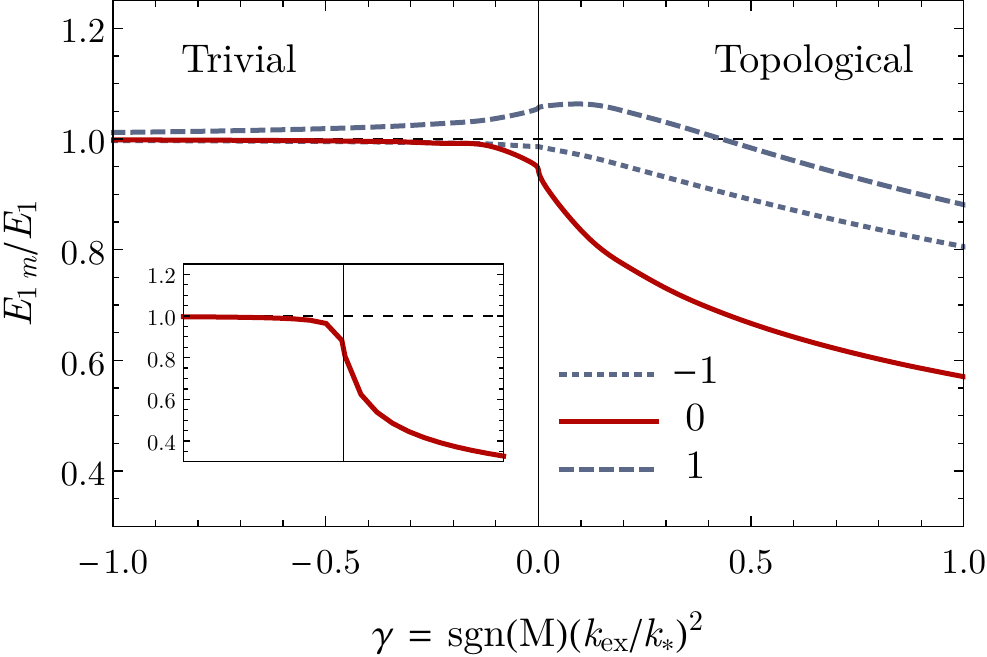}
\caption{\label{fig:energies} The $n=1$ energies for excitons formed from the particles in the upper BHZ block as a function of the dimensionless parameter $\gamma = \sgn(M) (k_\text{ex}/k_\ast)^2 \propto M$, scaled by the appropriate 2D hydrogen energy. The quantities $k_\ast$ and $k_\text{ex}$ are defined following Eq.~\ref{eq:Lz}. The sign of $\gamma$ determines the topological phase, as labeled. The inset shows the single $n=0$ state for the same range of $\gamma$. There is a clear qualitiative difference in the behavior of the energy levels on either side of the transition, with a crossover between them. In the topological phase there is a large splitting of states due to Berry physics that is absent in the trivial phase. Furthermore, the $m=0$ states change energy quickly above the topological transition. Energies are obtained using the effective fine structure constant $\alpha = 0.4$.}
\end{figure}

We find that these expectations are indeed true, with key features of our main results given in Figure \ref{fig:energies}; the hierarchy of exciton energy levels is drastically altered as one moves from the topologically trivial phase through the topological transition into the nontrivial phase.
Within the topological phase, states with opposite orbital angular momentum are well split from each other and the $m=0$ angular momentum state is pushed to a lower energy than the rest, all of which would be degenerate in the absence of Berry curvature.
In the trivial phase, however, all states are nearly degenerate, with the splitting decreasing the further one tunes away from the topological transition.
Though there is no sharp feature at the topological transition itself that distinguishes these two regimes, as one might expect from a topological effect, this behavior can nonetheless be explained as arising from effects intimately tied to topological character.

In Sec.\ref{sec:semiclassics} we begin by presenting an intuitive understanding of the physics at play in this system, considering a semiclassical model as well as an effective Hamiltonian for Dirac-like systems.
In Sec.\ref{sec:excitons} we formulate the full exciton problem and discuss further how topological effects will be manifested.
In Sec.\ref{sec:numerics} we discuss the methods used to numerically calulate the exciton spectra in the regimes of interest, and present our main numerical results. 

\section{Semiclassical approach}\label{sec:semiclassics}
Before presenting the full quantum mechanical exciton problem we discuss its semiclassical counterpart. 
This not only gives a clear physical picture of the role of the Berry phase, but also captures its effect on the electronic spectrum. 
The reason is that the semiclassical method applied to the usual 2D excitonic Coulomb problem reproduces the full spectrum exactly, and not just the structure of highly excited states. 
This remarkable result provides a fair assurance that our analysis here will provide useful insights into the problem at hand.


We start from the Lagrangian $L(\vec{r}_\mathrm{e},\vec{r}_\mathrm{h}, \vec{p}_\mathrm{e}, \vec{p}_\mathrm{h})$ for the dynamics of interacting electron and hole wave-packets~\cite{Sundaram1999,Chang2008,Xiao2010} given by 
\begin{equation}\label{eq:LagrangianFull}
L=\sum_{\alpha=\mathrm{e},\mathrm{h}}\left({\dot{\vec{r}}}_\alpha\cdot\vec{p}_\alpha + \dot{\vec{p}}_\alpha\cdot\vec{A}_{\alpha,\vec{p}_\alpha} - E^\alpha_{\vec{p}_\alpha}\right)-V(\vec{r}_\mathrm{e}-\vec{r}_\mathrm{h}).
\end{equation}
Here $\vec{r}_{\text{e}(\text{h})}$ is the location of the electron (hole) wave-packet and $\vec{p}_{\text{e}(\text{h})}$ is its momentum. 
We approximate their dispersions as quadratic in the vicinity of band minima as $E^\alpha_\vec{p}=\alpha \, \epsilon_p + \vec{p}^2/2 m$, where $\epsilon_p$ is the particle-hole asymmetry as defined after Eq.~\ref{eq:BHZham}. $V(\vec{r})=e^2/\epsilon r $ is the Coulomb interaction with dielectric constant $\epsilon$. 
The function $\vec{A}_{\alpha,\vec{p}_\alpha} = i\bra{\vec{p},\alpha}\boldsymbol{\nabla}_p\ket{\vec{p},\alpha}$ is the Berry connection, calculated from the particle and hole states of the BHZ Hamiltonian \ref{eq:BHZham}.
We consider only intrablock excitons with zero center of mass momentum $\vec{q}_\text{CM}=0$ since only they are optically active and are probed in experiments. 
The resulting Lagrangian for the relative motion of electron and hole is given by 
\begin{equation}\label{eq:LagrangianRel}
L=\dot{\vec{r}}\cdot\vec{p} + \dot{\vec{p}}\cdot \vec{A}_\vec{p} - \frac{\vec{p}^2}{2 \mu}-V(\vec{r}),
\end{equation}
where $\mu=m/2$ is the reduced electron-hole effective mass and the corresponding energy is independent of electron-hole asymmetry $\epsilon_\vec{p}$ of the BHZ model; 
$\vec{A}_\vec{p} = \vec{A}_{\text{e},\vec{p}}+\vec{A}_{\text{h},-\vec{p}}$ is the Berry connection for the relative electron-hole motion, which for the BHZ model is found to be (see Appendix \ref{sec:BHZBerry})
\begin{equation}\label{eq:BerryConnection}
\vec{A}_\vec{p} = -\left(s+\frac{M_\vec{p}}{\abs{\vec{d}_\vec{p}}}\right)\frac{\hat{\vec{z}}\times\vec{p}}{p^2},
\end{equation}
where $s\equiv\sgn{M/B}$ with the corresponding Berry curvature
\begin{equation} \label{eq:BerryCurvature}
\vec{\Omega}_\vec{p} = \boldsymbol{\nabla}_p \times \vec{A}_\vec{p} = A^2 \frac{M+Bp^2}{\abs{\vec{d}_\vec{p}}^3}\hat{\vec{z}}. 
\end{equation}
These two functions contain the topological information of two particle states within the model, with the integral of the Berry curvature over all momentum space giving the Chern number, a topological invariant that distinguishes topological and trivial phases. 
Note that the Berry connection is not gauge invariant, changing by the divergence of a scalar function if the state vectors are transformed by multiplication with a momentum dependent phase, but the Berry curvature is invariant under such transformations. 

The Euler-Lagrange equations obtained from (\ref{eq:LagrangianRel}) are given by
\begin{equation}\label{eq:EOM1}
\begin{gathered}
\dot{\vec{p}} = -\boldsymbol{\nabla}_r V(\vec{r}) \\
\dot{\vec{r}} = \frac{\vec{p}}{\mu} + \boldsymbol{\nabla}_r V(\vec{r})\times \vec{\Omega}_\vec{p}.
\end{gathered}
\end{equation}	
The term containing the electron-hole Berry curvature is the anomalous velocity.
Examining these equations in polar coordinates 
shows that the anomalous velocity contributes only to the angular motion of the exciton, with left-spinning and right-spinning states affected exactly oppositely, breaking the symmetry between them that is present in the absence of the Berry curvature.
In particular, the anomalous velocity changes the usual expression detailing the conservation of angular momentum to 
\begin{equation}\label{eq:Lz}
L_\text{z}=\left[\vec{r}\times\vec{p}\right]_z + \frac{\Phi_\vec{p}}{2\pi} \hat{\vec{z}}.
\end{equation}
Here $\Phi_\vec{p}$ is the Berry phase acquired by traversing a circular trajectory with momentum $\vec{p}$. 
From this observation one can anticipate that states with opposite angular momenta will not have the same energy in this system, unlike the case for the 2D hydrogen atom.

We can begin to get a more quantitative intuition for how topologically relevant physics comes into play through the function $\Phi_\vec{p}$.
To do so we compare two momentum scales. 
The first is the topological scale $k_\ast = \sqrt{\abs{M/B}}$. 
In the topological regime, since $M$ and $B$ have the same sign, the momentum-dependent Dirac mass $M_\vec{k} = M-Bk^2$ changes sign at this momentum, making it the most relevant momentum scale in the context of topological effects.
The Berry curvature reaches its maximum value near this momentum in the topological phase, and it is the point at which the function $\Phi_\vec{p}$ grows towards approximating the Chern number.  
In the trivial regime, though this is a well defined momentum, there are no effects of note at this scale. 
The second scale is the characteristic momentum for excitonic physics, related to the inverse Bohr radius of the exciton, $k_\text{ex} = 1/a_B = \mu e^2/\epsilon$.
In the topological phase, if the excitonic momentum is small compared to the topological scale, then the Berry phase term is likewise small.
This is always the case in the trivial phase since the Berry curvature is small for all reasonable momenta (see Appendix \ref{sec:BHZBerry}).
If the ratio of these two scales becomes even moderately sized in the topological phase, however, then the Berry phase will become a nontrivial perturbation to the angular momentum.
Though it is not immediately apparent how such a shift will affect the exciton spectrum, it is clear that any effect will only occur in the topological phase when $k_\text{ex}/k_\ast$ becomes sufficiently large. 

Additional insight can be made by changing coordinates, rewriting the angular momentum as $L_\text{z} =\left(\vec{R}\times \vec{p}\right)_z$. 
Here $\vec{R}=\vec{r}-\vec{A}_\vec{p}$ and $\vec{p}$ are the canonical coordinates of the problem, and the shift in the position coordinate is the momentum space equivalent of the Peierls substitution, which takes the Berry connection correctly into account. 
Using these coordinates the equations of motion (\ref{eq:EOM1}) can be derived from the effective Hamiltonian $H_\text{eff}(\vec{R},\vec{p})$ given by 
\begin{equation}
H_\text{eff}=\frac{\vec{p}^2}{2\mu}+V(\vec{R}+\vec{A}_\vec{p}).
\end{equation}
Expanding in the Berry connection and taking into account its solenoidal distribution in momentum space we get $H_\text{eff}=H_0+\Delta H$, where $H_0$ is the Hamiltonian of the usual Coulomb problem
\begin{equation}
H_0=\frac{P^2}{2\mu}+\frac{L_\mathrm{z}^2}{2\mu R^2} - \frac{e^2}{\epsilon R},
\end{equation}
and $\Delta H$ is the correction due to the presence of the Berry curvature and is given by  
\begin{equation}\label{DeltaH}
\Delta H = \left(s+\frac{M_{P}}{\abs{\vec{d}_{P}}}\right) \frac{e^2}{\epsilon R} \frac{L_\text{z}}{(R P)^2}.
\end{equation}
Here we introduced the canonical radial coordinates $R$ and $P$ along with the corresponding angular quantities $\phi$ and $L_\text{z}$. 
Examining this correction term we see that it acts as a perturbation to the 2D hydrogen atom problem and will generically split energy levels with differing angular momentum, as we determined to be the effect of the Berry phase above. 

The expression for $\Delta H$ we find here is a generalization of a correction derived previously using the Foldy-Wouthuysen transformation in the particular case of a constant Berry curvature \cite{Zhou2015, Trushin2017}.
Indeed, in the zero-momentum limit $\Delta H$ reduces to that form.
The Foldy-Wouthuysen transformation, however, generates another term that our semiclassical analysis is unable to reproduce--the Darwin term with the form $H_\text{Darwin} = \tfrac{1}{4}\Omega_z \nabla^2 V(R)$, where $\Omega_z = \Omega_z(\vec{p}=0)$.
This term gives an effective shift of the angular momentum by a value of $\tfrac{1}{2}$ in the perturbation to the hydrogen atom problem. 
This shift leads to an assymmetric splitting of states with opposite angular momentum as well as a shift for the $m=0$ state, which, in this effective Hamiltonian picture, would otherwise remain unaffected by Berry physics.

%

\section{Excitonic states}\label{sec:excitons}
Excitons are two-particle electron-hole bound states formed due to Coulomb interactions. 
Only excitons with zero total momentum $\vec{q}_\text{CM}=0$ are optically active and will be considered here.
Excitonic states can be written
\begin{equation} \label{eq:excitonstate}
\ket{\mathrm{X}_{i,j}} = \sum_{\vec{k}} C_{\vec{k}}^{(ij)} \,\, a^\dagger_{\vec{k},+,i}a_{\vec{k},-,j}\ket{0}.
\end{equation}
Here $C_{\vec{k}}^{(ij)}$ is the wave function of the exciton in momentum space, $\ket{0}$ is the state with filled valence bands and empty conduction bands, and $a^\dagger_{\vec{k},\alpha,i}$ ($a_{\vec{k},\alpha,i}$) creates (destroys) the single particle state $\ket{\vec{k},\alpha,i}$, where $\alpha=\pm$ labels the band and $i=1,2$ labels the block of the BHZ Hamiltonian the state is taken from. 
When rotated to the band basis, each of the two $2\times 2$ blocks of the BHZ Hamiltonian produce a single conduction and valence band, hosting the electrons and holes that are the building blocks of excitons.
When $i=j$ then the electron and hole come from the same block (intrablock excitons), while the case of $i\neq j$ corresponds to an interblock exciton.
In general intrablock excitons are optically active while interblock excitons require some degree of inversion symmetry breaking to be accessable via optical means.
We do not consider such symmetry breaking in our model, but we calculate interblock exciton energies nonetheless as a point of comparison.

The exciton wave function satisfies a Schrodinger-like equation in momentum space given by
\begin{equation}\label{eq:Schrodinger}
2 \abs{\vec{d}_\vec{k}} C_{\vec{k}}^{(ij)} - \sum_{\vec{k}'} U_{\vec{k}-\vec{k}'}\mathcal{F}^{(ij)}_{\vec{k},\vec{k}'}C_{\vec{k}'}^{(ij)} = (E_g + E_\text{X}) C_{\vec{k}}^{(ij)}.
\end{equation}
Here $E_g = 2\abs{M}$ is the energy gap and $E_\text{X}$ is the exciton energy.
The screened Coulomb interaction is given by $U_\vec{q}=2\pi e^2/\epsilon q$, with $\epsilon$ being the effective dielectric constant of surrounding medium, and $2\abs{\vec{d}_\vec{k}} = E_{k,+,i} - E_{k,-,j}$ is the two-particle free dispersion, independent of the particle-hole asymmetry.
Finally, $\mathcal{F}^{(ij)}$ is a function resulting from the rotation from the original basis of the Hamiltonian to the band basis, and its importance will be discussed at length.

In our further analysis we will approximate the two particle dispersion with a constant parabolic dispersion,
\begin{equation}
2\abs{\vec{d}_\vec{k}} \to 2\abs{M} + \frac{k^2}{2\mu}, \quad \text{with }\mu = \frac{\abs{M}}{2A^2}.
\end{equation}
This changes the features of the spectrum in the topological phase, which for the unmodified BHZ model develops a degenerate band minimum at a finite momentum for large enough $|M|$; in a more accurate approximation there is a value of $M$ for which the effective mass near $k=0$ changes sign within the topological phase.
This alone can lead to large effects on excitonic properties but is completely unrelated to the topological transition.
Since smooth deformations of the band structure leave topological properties unchanged, the above simplification is one way to remove this parametric dependence of the model on $M$ while leaving topological properties intact.
This allows us to more easily isolate the effect of topology alone. 

Since this model has rotational symmetry we can perform a full multipole decomposition, writing the exciton wave function as
\begin{equation}
C_{\vec{k}}=\sum_m C_m(k) e^{i m \varphi_\vec{k}}.
\end{equation}
The eigenvalue equation itself becomes 
\begin{equation}
\frac{k^2}{2\mu} C_m(k) - \sum_{k'} U^{\text{eff}}_{m}(k,k')C_m(k') = E_\text{X} C_m(k),
\end{equation}
with $U^\text{eff}_m$, the effective interaction in the $m$ channel, given by
\begin{equation}
U^\text{eff}_m(k,k')=\sum_{m'}U_{m-m'}(k,k') F_{m'}(k,k').
\end{equation}
Being the index related to rational invariance, $m$ is a component of the angular momentum of the exciton, specifically the component related to the relative motion of its constituents (see Appendix \ref{sec:symmetries} for the full angular momentum). 
It should be noted the choice of the underlying spinor wave functions of electrons and holes is not unique (see Appendix \ref{sec:BHZBerry}), and one can change them up to an arbitrary gauge transformation. 
Though gauge transformations leave all observables unchanged, they can in general uniformly shift the label $m$ by any integer, making this label of excitonic states ambiguous and dependent on gauge choice.
The gauge that we employ is chosen to reduce to the normal labeling of states for the 2D hydrogen atom in the limit $MB\to-\infty$, infinitely far into the trivial regime. 

\begin{figure}[!ht]
\center
\includegraphics[width=\columnwidth]{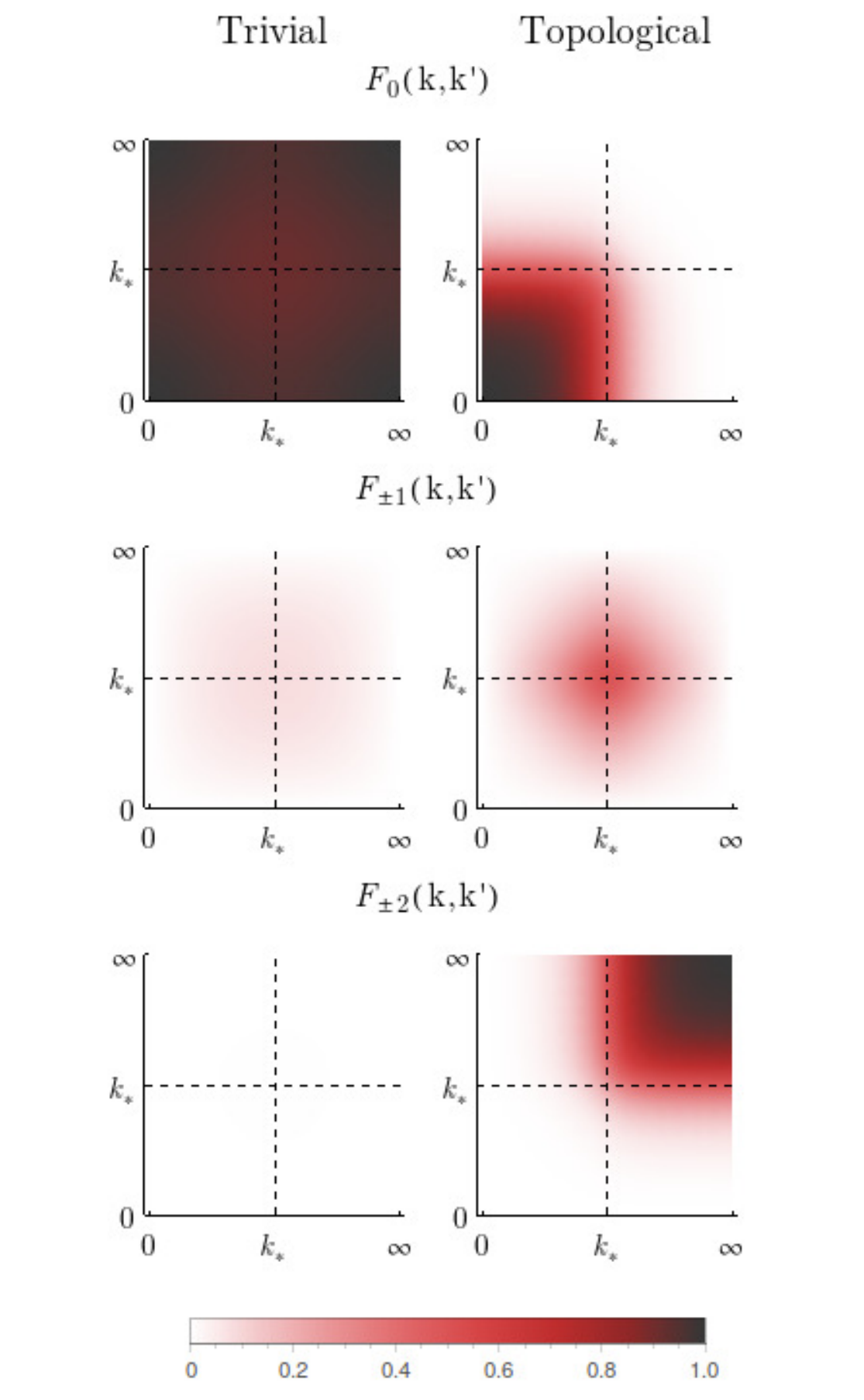}
\caption{\label{fig:overlap} The multipole coefficients of the overlap function $\mathcal{F}$ plotted as functions of their two arguments $k$ and $k'$, with the infinite domains $k,k'\in[0,\infty)$ projected onto a finite interval. The plots on the left show the typical behavior of the $F$'s in the trivial phase, and those on the right show them in the topologically nontrivial phase. There is a clear and distinct difference in the qualitative behavior of these functions on either side of the transition, with a sudden jump from one behavior to the other at the transition itself.}
\end{figure}

The function $\mathcal{F}^{(ij)}_{\vec{k},\vec{k}'}=\braket{\vec{k},+,i|\vec{k}',+,i}\braket{\vec{k}',-,j|\vec{k},-,j}$ results from the change to the band basis and is given by the overlaps of electron and hole spinor wave functions.
We can explicitly write this function as
\begin{equation}
\begin{split}
\mathcal{F}^{(11)}_{\vec{k},\vec{k}'} = e^{i(s-1)(\varphi_\vec{k}-\varphi_\vec{k'})}\cos^2\tfrac{\theta_\vec{k}}{2}\cos^2\tfrac{\theta_\vec{k'}}{2}\\
+ e^{i(s+1)(\varphi_\vec{k}-\varphi_\vec{k'})}\sin^2\tfrac{\theta_\vec{k}}{2}\sin^2\tfrac{\theta_\vec{k'}}{2}\\
+ 2e^{i s(\varphi_\vec{k}-\varphi_\vec{k'})}\cos\tfrac{\theta_\vec{k}}{2}\cos\tfrac{\theta_\vec{k'}}{2}\sin\tfrac{\theta_\vec{k}}{2}\sin\tfrac{\theta_\vec{k'}}{2},\\
\mathcal{F}^{(12)}_{\vec{k},\vec{k}'} = \cos^2\tfrac{\theta_\vec{k}}{2}\cos^2\tfrac{\theta_\vec{k'}}{2} + \sin^2\tfrac{\theta_\vec{k}}{2}\sin^2\tfrac{\theta_\vec{k'}}{2}\\
+ 2 \cos\tfrac{\theta_\vec{k}}{2}\cos\tfrac{\theta_\vec{k'}}{2}\sin\tfrac{\theta_\vec{k}}{2}\sin\tfrac{\theta_\vec{k'}}{2} \cos(\varphi_\vec{k}-\varphi_\vec{k'}),
\end{split}
\end{equation}
with $\mathcal{F}^{(11)} = \mathcal{F}^{(22)\ast}$, $\mathcal{F}^{(12)}=\mathcal{F}^{(21)}$, $\cos\theta_\vec{k} = M_\vec{k}/\abs{\vec{d}_\vec{k}}$ and $s\equiv\sgn M$.
This is the only ingredient in the excitonic eigenvalue equation~(\ref{eq:Schrodinger}) which reflects the underlying topology, and it is qualitatively different in trivial and topological regimes.
The topological information carried in these functions can be seen explicitly by considering their multipole expansions,
\begin{equation}
\begin{gathered}
\mathcal{F}^{(11)}_{\vec{k},\vec{k}'} = \sum_{m=0}^2 F_{sm}(k,k') e^{ism(\varphi_\vec{k}-\varphi_\vec{k'})},\\
\mathcal{F}^{(12)}_{\vec{k},\vec{k}'} = F_0(k,k') + F_{2s}(k,k') + F_{s}(k,k')\cos(\varphi_\vec{k}-\varphi_\vec{k'}),
\end{gathered}
\end{equation}
and examining the behavior of the three functions $F_{sm}$ in both the trivial and topological phases.
These functions are plotted in Figure~\ref{fig:overlap}.

Far into the trivial parameter regime, one sees that $F_0$ is approximately equal to $1$ for all values of $k,k'$, while the other functions are very small.
Indeed in the limit $MB\to-\infty$ then $F_0\to 1$ and $F_{m\neq 0}\to 0$, so $\mathcal{F}\to 1$ and the Schrodinger equation approaches exactly that for the 2D hydrogen atom. 

A similar statement cannot be made in the topological parameter regime, with the behavior of $\mathcal{F}$ being fundamentally nontrivial for all values of the tuning parameter. 
In this regime both $F_0$ and $F_{-2}$ show nontrivial behavior as one or both of their arguments become large compared to the topological scale $k_\ast$.
The remaining function, $F_{-1}$ does not display such a drastic change on either side of the topological transition, though in the topological phase it does always reach the value $1/2$ for $k=k'=k_\ast$. 
Note that these differences in the qualitative behavior of these functions are indeed tied directly to the topological character of the respective phases.
There is a sudden transition between one behavior and the other as the tuning parameter passes through the topological transition, with the $(k,k') \to(\infty,\infty)$ limits of the functions $F_0$ and $F_{\pm 2}$ changing discontinuously at that point. 

Both to gain further physical insight and to simplify eventual numerical integration, we rewrite the eigenvalue problem in a dimensionless form by scaling all momenta by the characteristic exciton momentum, $k_\text{ex} = 1/a_B = \mu e^2/\epsilon$.
Since the Bohr radius is the most natural length scale in the problem, its inverse gives the most relevant momentum for excitonic physics. 
This rescaling naturally results in an equation with only two dimensionless parameters: the relative fine structure constant of the material $\alpha = e^2/\epsilon A$ and the quantity $\gamma = (k_\text{ex}/k_\ast)^2$, comparing the size of the excitonic and topological momentum scales, which we use as our tuning parameter.

As we first considered in Sec.\ref{sec:semiclassics}, we can understand how the topological nature of the system manifests itself in excitonic properties by considering the relative size of these momentum scales, i.e.\ the size of $\gamma$. 
First note that the small momentum features of the functions in Figure \ref{fig:overlap} are very much alike on either side of the transition.
Close to the transition, where $|\gamma|$ is small and small momenta are most important, then excitons in both sides of the topological transition should be qualitatively similar.
Conversely, if $|\gamma|$ is not small then the nontrivial features of the $F$ functions near $k\sim k_\ast$ will be relevant in the topological phase and excitons should behave quite differently depending on the sign of $\gamma$.
From this we anticipate that our numerical analysis will not find a sharp feature in excitonic properties at the transition itself, instead seeing a gradual crossover between two behavior regimes. 


Another way to see the effects of topology is to note how this exciton problem compares with the two-dimensional hydrogen atom. 
Just as for the 2D hydrogen atom, the eigenstates of the exciton problem in this model are labeled by two indices, $n=0,1,2,\dots$ and $m=0,\pm 1,\dots, \pm n$, the principal and angular momentum quantum numbers.
\footnote{Another common choice of quantum numbers for the 2D problem is $n_r = 0,1,\dots$ and $m=0,\pm 1, \pm 2,\dots$, related to our choice by $n = n_r + \abs{m}$.
For our purposes $n$ and $m$ as defined in the text will prove more convenient.}
For the 2D hydrogen atom a hidden SO(3) symmetry (distinct from, but containing SO(2) rotational symmetry) ensures a perfect degeneracy between the $2n+1$ angular momentum states for each $n$ \cite{Yang1991, Parfitt2002}, with energies given by
\begin{equation}\label{eq:hydrogen}
E_{0,n} = - \frac{\mu e^4}{2\epsilon^2}\frac{1}{\left(n+\frac{1}{2}\right)^2}.
\end{equation}
In our system, though $n$ and $m$ are still good quantum numbers, the nontrivial overlap function $\mathcal{F}$ breaks the SO(3) symmetry, mixing angular momentum channels of the Coulomb interaction and reorganizing the spectrum.
The result is that the different angular momentum states for each energy level $n$ will have their energies split from each other, as anticipated in Sec.\ref{sec:semiclassics}.
Note that for interblock excitons states with angular momentum differing by a sign must still be degenerate due to time reversal symmetry, but those with different values of $|m|$ will generically be split. 

In general the breaking of this symmetry is ensured by the existence of any nonzero Berry curvature ($\mathcal{F}\neq 1$), even in a phase with trivial Chern number.
However, far enough into the topologically trivial phase one can consider this symmetry breaking as just a small perturbation to the 2D hydrogen atom problem (i.e. $\mathcal{F} \approx 1 + \delta\mathcal{F}$ with $\delta\mathcal{F}\ll 1$), which only introduces a small splitting between the states. 
The same cannot be said of the topologically nontrivial phase, where the behavior of $\mathcal{F}$ is fundamentally nontrivial as well, as described above and in Fig.\ref{fig:overlap}.
For this case the effect cannot be approximated as a small perturbation to the 2D hydrogen atom, so we can expect that the splitting between states will not necessarily be vanishingly small.



\section{Numerical Analysis}\label{sec:numerics}
We discretize the momentum in the integral Schrodinger equation according to a modified Gaussian quadrature method (with $N=192$ points) that is designed to handle the divergence in the Coulomb potential at $\vec k = \vec{k}'$ \cite{Chao1991}. 
Choosing a set value for $\alpha$ we can then invert the resulting matrix equation to find the excitonic spectrum as a function of the parameter $\gamma$.

In addition to the results presented in Fig.~\ref{fig:energies} showing the $n=0$ state and three $n=1$ states for intrablock excitons, we also calculated the corresponding states for interblock excitons, finding a similar effect.
The qualitative behavior of the exciton energy levels in the topologically trivial and nontrivial regimes is immediately apparent, with the different angular momentum levels separating from each other quickly as a function of $\gamma$ in the nontrivial regime, and converging to the 2D hydrogen energy moving deeper into the normal regime, as expected based on the discussed properties of the function $\mathcal{F}$.
Furthermore we find that intrablock exciton levels with opposite angular momentum split from each other, while corresponding levels in interblock excitons remain generate as ensured by symmetry.
We also note that while there is a crossover between two behavior regimes there is no sharp feature at the topological transition itself, again as anticipated.

The most notable behavior, seen in all cases, is the strong dependence of the $m=0$ state on $\gamma$, which has considerably smaller energy in the topological regime compared to the trivial regime. 
Indeed, for interblock excitons this is the primary feature we find.
Since this effect is found in all cases, it must be caused by a different mechanism than that causing the splitting of opposite angular momentum states in the intrablock case, i.e. the Berry phase. 
In other words, it is an effect that is insensitive to time reversal and seems to be strongest for the cases of zero angular momentum.

\section{Conclusions}
By examining excitonic spectra in the bulk of a model with nontrivial topology we have demonstrated that topology can have strong manifestations in bulk physics. 
In particular we have shown that the degeneracy of 2D excitonic states that would exist in a system without a Berry curvature is broken in the BHZ model due to the inclusion of such physical effects. 
In the trivial phase with Chern number 0 the splitting is small, with the Berry phase acting as a small perturbation to the 2D hydrogen atom problem. 
On the other side of the topological phase transition, however, the splitting is much greater since the effects of nontrivial topology can no longer be considered as just a small perturbation.
Though there is no sharp feature precisely at the transition point the difference in the behavior in the two phases can nevertheless be understood as a result of a change in the topological character.
As the characteristic excitonic momentum scale becomes sizable compared to the scale associated with topological effects then the large momentum differences between the physics of the trivial and nontrivial phases becomes essential.
The result is a dramatic reorganization of the excitonic spectrum, producing a hierarchy of states that is utterly distinct for values of the tuning parameter well into each of the two phases.

\section{Acknowledgements}
This work was supported by NSF-DMR 1613029 (A.A.A.), the DOE-BES (DESC0001911) and the Simons Foundation (V.M.G.). 
We would like to thank Mehdi Kargarian for helpful discussions.

\appendix
\section{Symmetries and Angular Momentum}\label{sec:symmetries}
BHZ Hamiltonian respects both time-reversal and inversion symmetry, with the two blocks of the Hamiltonian (\ref{eq:BHZham}) mapping into each other under time reversal and remaining unchanged with inversion. 
This can be explicity verified by representing the time reversal and inversion operators respectively as
\begin{equation}
\Theta = -i\hat\sigma_y K\otimes\hat{\mathbf{1}}, \text{ and } P = \hat{\vec{1}}\otimes\hat\tau_z,
\end{equation} 
and confirming that they commute with the Hamiltonian.
Here $K$ denotes complex conjugation. 
The set of single particle eigenstates respects these symmetries as well, which one can straightforwardly verify, finding
\begin{equation}
\begin{gathered}
\Theta \ket{\vec{k},\pm,i} = \sum_j\epsilon_{ij}\ket{-\vec{k},\pm,j}\\
P\ket{\vec{k},\pm,i} = \pm\ket{-\vec{k},\pm,i}.
\end{gathered}
\end{equation}
Here the $\pm$ labels the conduction and valence bands, while $i,j$ labels the block of the Hamiltonian that acts on the states.

In addition to these discrete symmetries the system is also rotationally invariant so total angular momentum is also a good quantum number. 
Since the system is two-dimensional, the total angular momentum is equivalent to its z-component.
The angular momentum of a particles has three components---spin, $S_z=\hat{\vec{1}}\otimes\hat{\sigma}_z/2$, atomic orbital, $K_z = \text{diag}(0,1,0,-1)$, and orbital, $L_z = \hat{\mathbf{1}} \left(\vec{r}\times\vec{p}\right)_z$---so in total we have $J_z = S_z + K_z + L_z$, and a simple calculation confirms that $\left[J_z,H_\text{BHZ}\right]=0$.

The eigenstates of $H_\text{BHZ}$ are also eigenstates of $J_z$, and we can most easily compute the angular momentum of single particle states at $\vec{k}=0$, though the result must hold at all points in k-space, giving
\begin{equation}\label{eq:singlejz}
J_z\ket{\vec{k},\pm,i} = (-1)^{i+1}\left[1 \mp \tfrac{1}{2}\sgn M\right]\ket{\vec{k},\pm,i}.
\end{equation}
Note that this value is simply either $\tfrac{1}{2}$ or $\tfrac{3}{2}$ up to a sign.

To consider excitons we must we add the Coulomb interaction to this single particle Hamiltonian.
With regards to symmetry it is enough to note that the Coulomb interaction is also invariant under time reversal, inversion, and rotations, so the states of the interacting system must obey these symmetries as well.
Let intrablock exciton eigenstates be labeled as $\ket{n,m,i}$, where $n$ and $m$ are two quantum numbers, and $i$ labels the block we take the particle and hole contituents from. 
In the center of mass frame these exciton eigenstates are
\begin{equation} 
\ket{n,m,i} = \sum_\vec{k} C_{nm\vec{k}}^{(i)}\,a^\dagger_{\vec{k},+,i}a_{\vec{k},-,i}\ket{0}.
\end{equation}

A straightforward calculation shows that
\begin{equation} \label{eq:excitonjz}
J_z\ket{n,m,i} = \left[m + (-1)^i\sgn M\right]\ket{n,m,i} \equiv j\ket{n,m,i},
\end{equation}
where the second term in the eigenvalue is the sum of the spin and orbital angular momenta of the single particle bands, Eq.\ref{eq:singlejz}.
We see here that $m$ labels the part of the angular momentum interpreted classically as arising from the relative motion of the exciton's constituent particle and hole.
Furthermore, it can be easily verified that time reversal acts in the expected way, simply flipping the sign of the angular momentum, $
J_z \Theta \ket{n,m,i} = -j\, \Theta \ket{n,m,i}$.

\section{Berry Physics in BHZ model} \label{sec:BHZBerry}
The spinor eigenstates corresponding to the two bands of the upper block of the Hamiltonian (\ref{eq:BHZham}) are
\begin{equation} \label{eq:states}
\begin{gathered}
\ket{\vec{k},+} = \begin{pmatrix}
e^{i\frac{s+1}{2}\varphi_\vec{k}} \cos\frac{\theta_\vec{k}}{2} \\
e^{i\frac{s-1}{2}\varphi_\vec{k}} \sin\frac{\theta_\vec{k}}{2},
\end{pmatrix},\\
\ket{\vec{k},-} = \begin{pmatrix}
-e^{-i\frac{s-1}{2}\varphi_\vec{k}} \sin\frac{\theta_\vec{k}}{2} \\
e^{-i\frac{s+1}{2}\varphi_\vec{k}} \cos\frac{\theta_\vec{k}}{2},
\end{pmatrix}
\end{gathered}
\end{equation}
where $s \equiv \sgn MB$ and $\cos\theta_\vec{k} =  M_\vec{k}/\abs{\vec{d}_\vec{k}}$. 
The states for the other block can be generated from these by applying the time reversal operator, discussed in Appendix \ref{sec:symmetries}.

\begin{figure}
\center
\includegraphics[width=0.9\columnwidth]{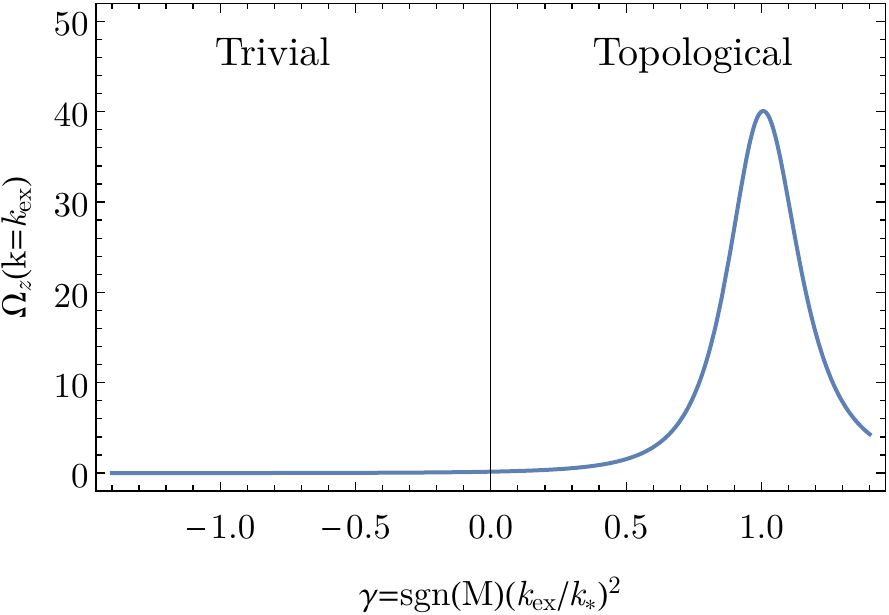}
\caption{\label{fig:BerryCurvature} A plot of the z-component of the Berry curvature $\vec{\Omega_+(\vec{k})}$ as a function of momentum in both the topological (top) and trivial (bottom) phases. The Berry curvature $\vec{\Omega_-(\vec{k})}$ is simply related by a sign. The momentum is measured in units of the topological scale $|k_\ast| = \sqrt{|M/B|}$. In the topological regime the Berry curvature is peaked near $k_\ast$ and is positive for all values of the momentum, leading to a nonzero Chern number, while in the trivial regime it takes both positive and negative values producing a Chern number of 0.}
\end{figure}

The information about the topology is stored in the Berry connection of electrons defined in terms of these states as
\begin{equation} \label{eq:1pBerryConnection}
\vec{A}_\pm(\vec{k}) = i \bra{\vec{k},\pm} \boldsymbol{\nabla}_k \ket{\vec{k},\pm} = \mp\frac{s+\cos\theta_\vec{k}}{2k^2} \left(\hat{\vec{z}}\times\vec{k}\right).
\end{equation}
The corresponding Berry curvature for the upper block is
\begin{equation}
\vec{\Omega}_\pm(\vec{k}) = \boldsymbol{\nabla}_k \times A_\pm(\vec{k}) = \pm A^2 \frac{M+Bk^2}{2d(k)^3}\hat{\vec{z}}.
\end{equation}
It is gauge independent, and its integral gives the Chern number distinguishing topological phases. 
The momentum distribution of the the Berry curvature in the topological and trivial regimes is presented in Fig. \ref{fig:BerryCurvature}.
Calculating the same quantities for the lower block of the Hamiltonian gives the same results up to overall signs. 

When considering the interacting two particle problem we define the particle and hole states as
\begin{equation}
\ket{\vec{k},e} = \ket{\vec{k},+},\quad \ket{\vec{k},h} = \mathcal{C}\ket{\vec{k},-},
\end{equation}
using the particle-hole transformation $\mathcal{C} = K\hat{\sigma}_x$.
With these definitions we can define the Berry connections for particles and holes in the upper block analogously as in Eq.~\ref{eq:1pBerryConnection} to find $\vec{A}_e(\vec{k}) = \vec{A}_+(\vec{k})$ and $\vec{A}_h(\vec{k}) = \vec{A}_-(\vec{k}) = -\vec{A}_e(\vec{k})$.

\bibliography{references}
\end{document}